\documentstyle[psfig]{lamuphys}
\makeatletter
\let\chapter\hid@chapter
\makeatother
\newcommand{\R}{${\cal R}$} 
\newcommand{\etal}{{\it et al.\ }}
\newcommand{\lta}{\stackrel{<}{\scriptstyle\sim}}
\newcommand{\gta}{\stackrel{>}{\scriptstyle\sim}}
\begin{document}
\pagenumbering{arabic}
\title{Predicting Spectral Properties of DLA Galaxies}

\author{Uta\,Fritze -- v. Alvensleben, Ulrich\,Lindner, Claudia S.\,M\"oller}

\institute{Universit\"ats-Sternwarte, Geismarlandstra{\ss}e 11,
D-37083 G\"ottingen, Germany}

\maketitle

\begin{abstract}
Comparison of our chemically consistent models for spiral galaxies 
with observed DLA abundances shows that at high redshift DLA 
galaxies may well be the progenitors of normal spiral disks of all 
types from Sa through Sd. Towards lower redshifts ${\rm z \leq 1.5}$ 
however, early type spirals drop out of DLA samples due to low gas 
or/and high dust content. We use the spectrophotometric aspects of our unified 
spectral, chemical, and cosmological evolution models to predict 
expected luminosities in different bands for DLA galaxies at various 
redshifts and compare to the few optical identifications available. 
\end{abstract}
\section{Introduction}
Within the framework of our unified spectrophotometric, chemical and 
cosmological modelling of galaxies of various types we use the Star Formation 
Histories ({\bf SFHs}) that gave satisfactory agreement of global 
galaxy colors, emission and absorption features with observations of galaxies 
nearby and up to high redshifts to study the chemical abundance evolution they 
imply. These models define various galaxy 
types in terms of their respective SFHs immediately give -- without any additional parameters -- 
good agreement, after a Hubble time, with HII region 
abundances in nearby galaxies. We then investigate the redshift evolution of a 
number of individual element abundances in the ISM of our spiral galaxy models 
and compare them to observed DLA abundances. From this comparison we derive our 
predictions as to the galaxy types that give rise to DLA absorption as well as 
for their spectrophotometric properties, both at low and high redshift. 

\section{Chemially Consistent Chemical, Spectral and Cosmological Galaxy Evolution Models}
Our unified chemical and spectrophotometric evolutionary synthesis models describe 
the spectrophotometric evolution in terms of spectra (UV -- NIR), luminositis, 
and colours and -- at the same time --  the chemical evoltion of a number of ISM element 
abundances, including SNIa contributions, as a function of time. 
When combined with a cosmological model specified by the parameters ${\rm H_0, \Omega_0, 
\Lambda_0}$ and a redshift of galaxy formation ${\rm z_{form}}$, we obtain all 
quantities as a function of redshift, in case of spectrophotometric properties 
including evolutionary and cosmological corrections as well as the effect of 
attenuation by intergalactic hydrogen randomly distributed along the lines of sight to very 
distant objects (Madau 1995). 

Our models are {\bf chemically consistent} in the sense that we keep track of ISM abundances 
at birth of each star and account for the increasing initial metallicity of successive 
generations of stars by using different sets of input physics, i.e. 
stellar evolutionary tracks, stellar spectra, colour calibrations, yields, lifetimes, 
and remnant masses, for a range of metallicities \ 
-2.3 $\leq$ [Fe/H] $\leq$ +0.3. The models have a strong analytic power 
and directly show the 
luminosity contributions to any wavelength band and the enrichment 
contributions to any chemical element due to different 
stellar masses, spectral types, luminosity classes, metallicity 
subpopulations, and nucleosynthetic origins (PNe, SNI, SNII, 
single stars, binaries, ...), and all this as a function of time 
or redshift. On the other hand, they are simple 
1-zone closed box models without any spatial resolution or dynamics 
included. While the ISM is assumed to be instantaneously and ideally 
mixed at any time, the finite lifetime of the stars before they give back 
enriched material to the ISM is fully taken into account. 

We use stellar evolutionary tracks from the Padova Group 
(Bressan \etal 1993, Fagotto \etal 1994a, b, c) 
for stars in the mass range 
0.6 ${\rm \leq  m_{\ast} \leq 120~M_{\odot}}$, and from 
Chabrier \& Baraffe (1997) for m${\rm _{\ast} \leq 0.5~M_{\odot}}$, 
stellar model atmosphere spectra from Lejeune \etal (1998), 
and stellar yields from v. d. Hoek \& Groenewegen (1997) for ${\rm m_{\ast} 
\leq 8~M_{\odot}}$, from 
Woosley \& Weaver (1995) for stars with ${\rm 12 \leq m_{\ast} \leq 40~M_{\odot}}$, 
and from Nomoto \etal (1997) for SNIa (cf. Lindner \etal 1999).  
We caution that the metallicity dependance of these yields, of the explosion energies, 
lifetimes, mass loss rates, etc. are 
not fully understood at present. \\
{\bf Parameters} in our models are the IMF and the SFH. We use a Scalo IMF and SFHs 
appropriate for the different galaxy types. For spirals Sa, ..., Sc the SFR is a 
linear function of the gas-to-total-mass ratio $\Psi(t) \sim \frac{G}{M} (t)$, while for our Sd 
model we adopt a constant SFR. Constants are chosen as to 
result in characteristic timescales for SF ${\rm t_{\ast}}$ (with ${\rm t_{\ast}}$ defined via 
${\rm \int_0^{t_{\ast}} 
\Psi \cdot dt = 0.63 \cdot G (t=0)}$) of 2, 3, 10, and 16 Gyr for galaxy types Sa, Sb, Sc, 
Sd, respectively. 

These SFHs provide agreement with nearby galaxy samples and templates of the respective 
spiral types in terms of colours (RC3), template spectra (Kennicutt 1992), and characteristic 
HII region abundances, i.e. HII region abundances as observed at ${\rm \sim 1~R_e}$ 
(Zaritsky \etal 1994, Oey \& Kennicutt 1993, Ferguson \etal 1998). Those are the ones to be 
compared to our 1-zone models as well as to DLA abundances (Phillipps \& Edmunds 1996). 

\section{Comparison with observed DLA Abundances}
High resolution spectroscopy has provided precise element abundances 
for C, N, O, Al, Si, S, Cr, Mn, Fe, Ni, Zn, ... 
for a large number of DLAs over the redshift range 
from z $\sim$ 0.4 through z $\gta$ 4.4 (see e.g. Pettini {\sl this conf.}, 
Boiss\'e \etal 1998, Lu \etal 1993, 1996, Pettini \etal 1994, 1998, 
Prochaska \& Wolfe 1997, de la Varga \& Reimers {\sl this conf.}). 
We have carefully referred all abundances to one homogeneous set of 
oscillator strengths and solar reference values. 

A comparison of the redshift evolution of our models with observed DLA abundances 
is presented in Fig. 1a. for the example of [Zn/H] which is not affected by depletion 
onto dust grains, and in Fig. 1b. for [Fe/H] which, under local conditions, is significantly 
depleted. For an extensive comparison including many elements as well as for a detailed 
description of our models see Lindner \etal (1999).

\begin{figure}
\vspace{-6pt}
\centerline{\psfig{figure=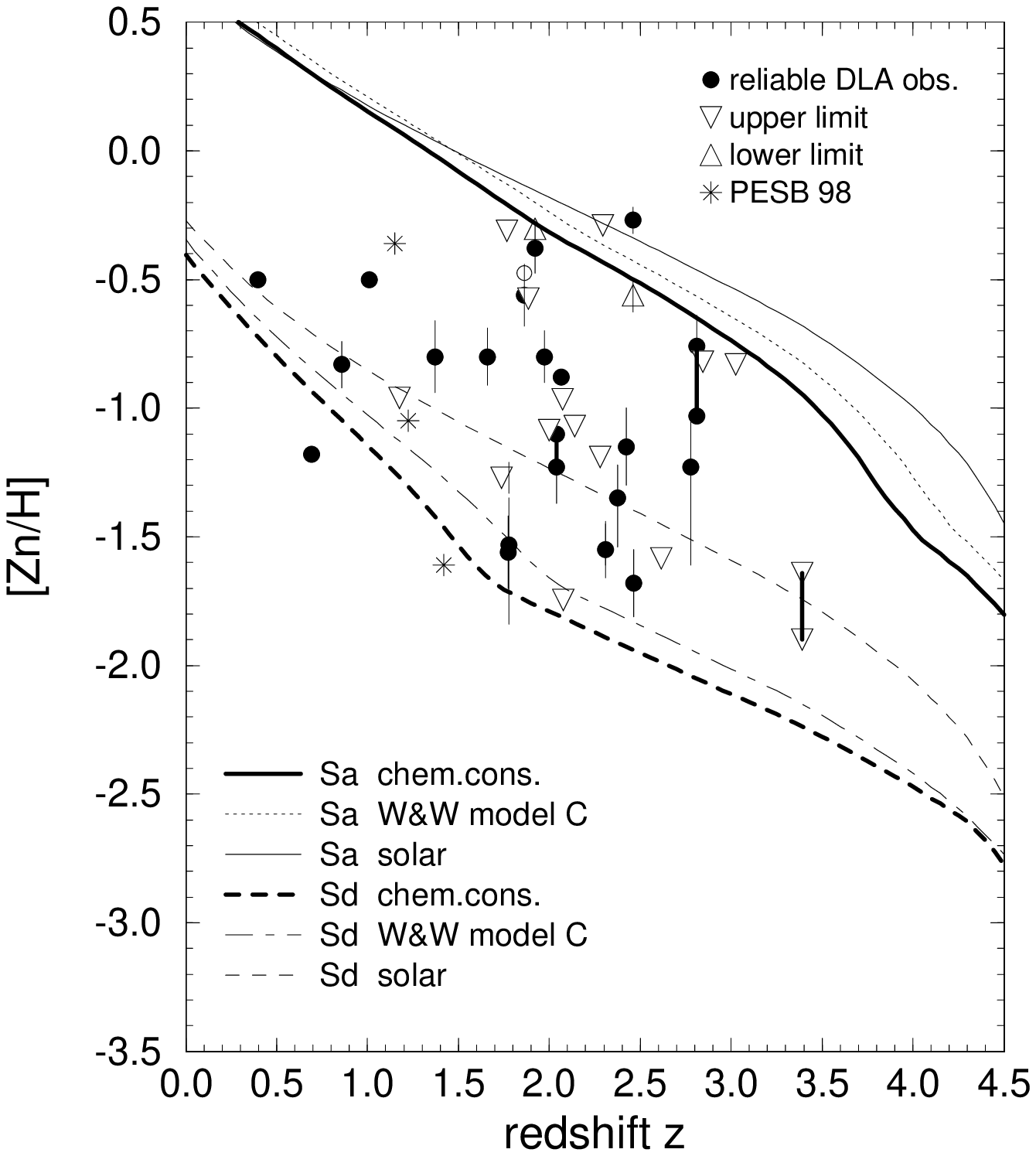,height=7cm}\psfig{figure=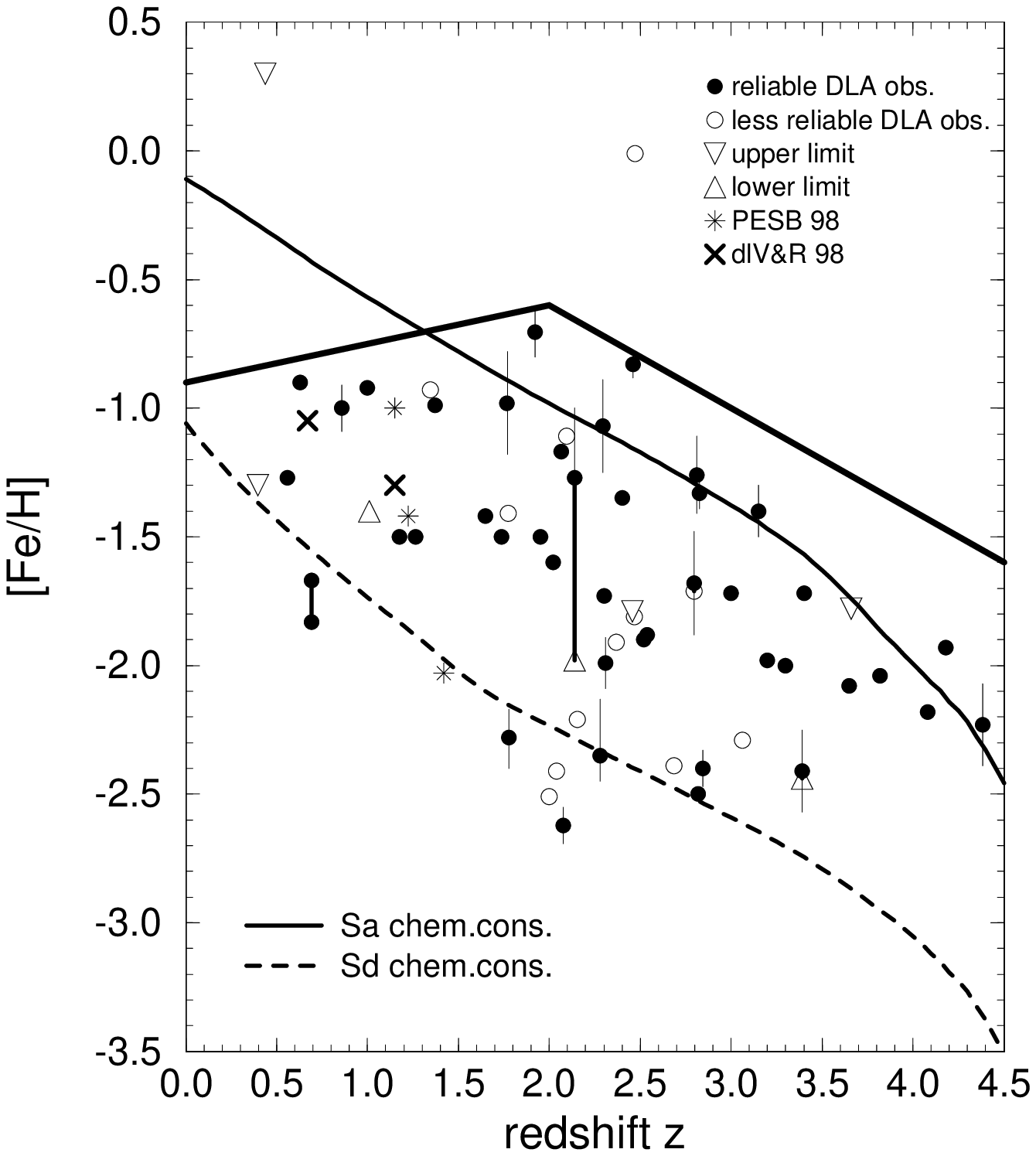,height=7cm}} 
\vspace{-5pt}
\caption{Redshift evolution of [Zn/H] {\bf (1a)} and of [Fe/H] {\bf (1b)}. 
Heavy lines are for our chemically consistent models, this lines are for models 
using solar metallicity input physics only. 
For clarity, only our Sa and Sd models are depicted, the evolution of Sb and Sc models falls 
between the latter.}
\vspace{-6pt}
\end{figure}

\noindent
The comparison for all elements available yields the following results: 
{\bf Sa -- Sd models bracket redshift evolution of DLA abundances from ${\rm \bf
z \gta 4.4}$ to ${\rm \bf z \sim 0.4}$, models bridge the gap from high-z DLAs  to 
nearby spiral HII region abundances, the weak redshift evolution of DLA abundances 
is a natural consequence of the long SF timescales for spiral galaxies, and the 
range of SF timescales 
for near-by spirals from Sa through Sd fully explains the abundance 
scatter among DLAs at fixed redshift.} 

We thus conclude that from the point of view of their chemical abundance history 
{\bf DLA galaxies might well be the progenitors of normal spirals}, 
allthough we cannot some starbursting dwarfs or giant 
LSB galaxies among the DLA galaxy population (see also Lindner \etal {\sl
this conf.}, Lindner \etal 1999). 

Beyond the overall agreement between abundances given by our spiral galaxy models and 
by high resolution DLA data, inspection of Figs 1a, b reveals that while at high 
redshift, DLA data fill all the range between our {\sl rapid evolution} Sa and {\sl 
slow evolution} Sd models, this seems no longer to be the case at low redshift. 
Below ${\rm z \lta 1}$ data points tend to fall close to our late type models. 
This is elucidated by the straight line in Fig. 1b which -- at the same time -- marks 
the 50\% global gas-to-total mass ratio in our models. The number of DLAs at ${\rm z \lta 1}$, 
however, is still quite small. 
If confirmed by further DLA detections at low redshift, this would indicate that the 
composition of the DLA galaxy population changes with redshift. 
{\bf While at high redshift the progenitors of all spiral types from Sa through Sd give rise to 
DLA absorption, the early type spirals drop out of the DLA sample towards lower redshift 
due to too scarce gas content} or/and too high dust content. Indeed, low-z 
DLAs are observed to have low hydrogen colunm densities as compared to their high-z counterparts 
(Meyer \etal 1995). 
This has important 
consequences for the prospects of optically identifying the galaxies responsible for 
DLA absorption. 

\section{Predictions for Spectral Properties of DLA Galaxies}
Average luminosities of early spiral types Sa (${\rm \langle M_B \rangle = -19.7 \pm 1.2}$) 
are brighter than those of late types Sd (${\rm \langle M_B \rangle = -17.7 \pm 1.5}$), 
allthough their luminosity ranges slightly overlap (e.g. 
Sandage \& Tammann 1985). On average, spirals are significantly 
fainter than ${\rm L^{\ast}}$. 

We can now use the spectrophotometric results from our models to predict average 
apparent luminosities 
in various passbands for our model galaxies as a function of redshift, including evolutionary 
and cosmological corrections and attenuation (cf. M\"oller, F.-v.A., Fricke 1998) and find 
the intriguing result that 
early type spirals in the redshift range ${\rm z \sim 2 - 3}$ have about the same 
apparent magnitudes in all 3 bands B, \R, K as the intrinsically fainter late type 
spirals Sd at the lowest 
redshifts ${\rm z \sim 0.5}$ where DLA absorption is seen.

\begin{figure}
\vspace{-20pt}
\centerline
{\psfig{figure=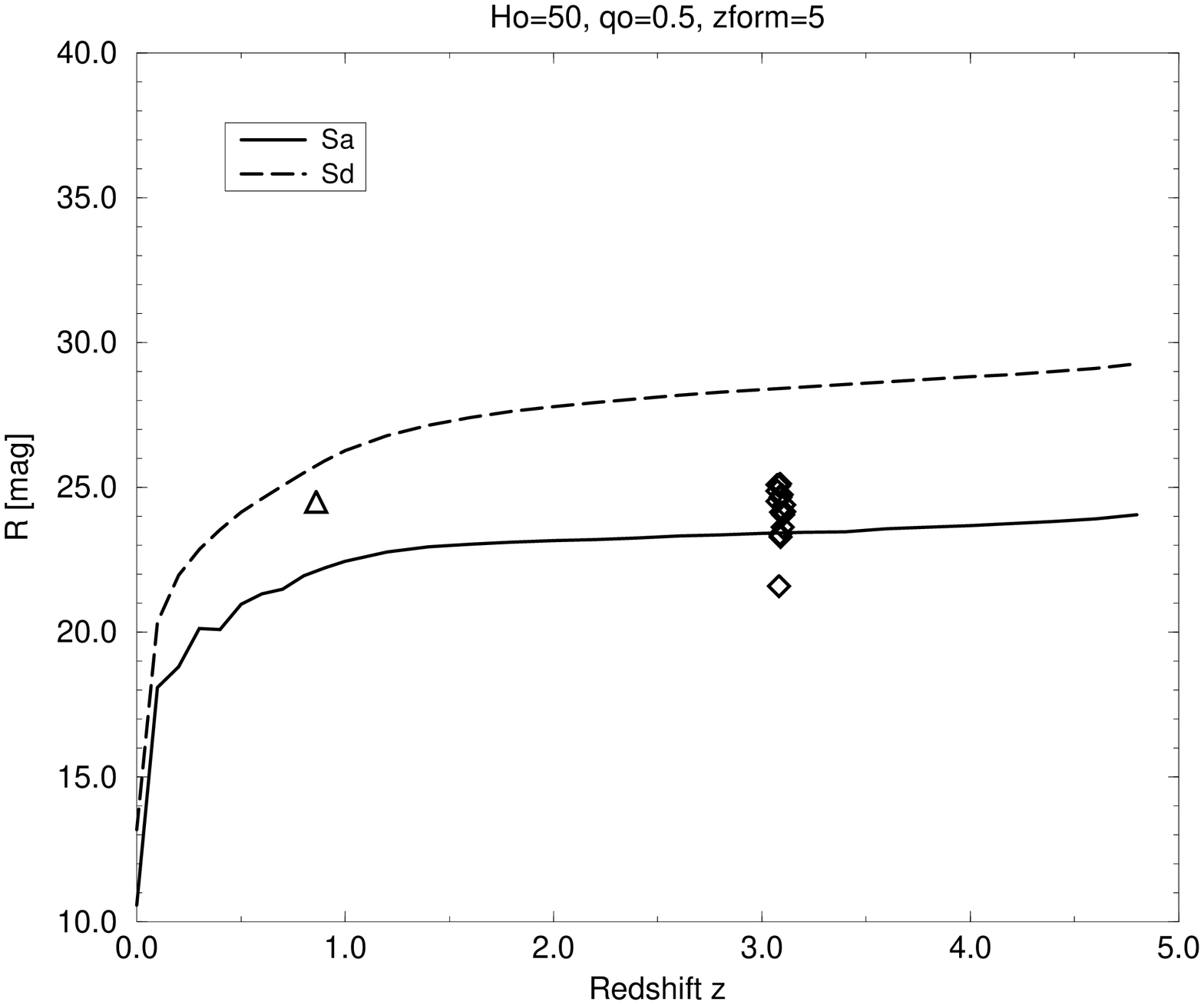,height=5.4cm,angle=270}\psfig{figure=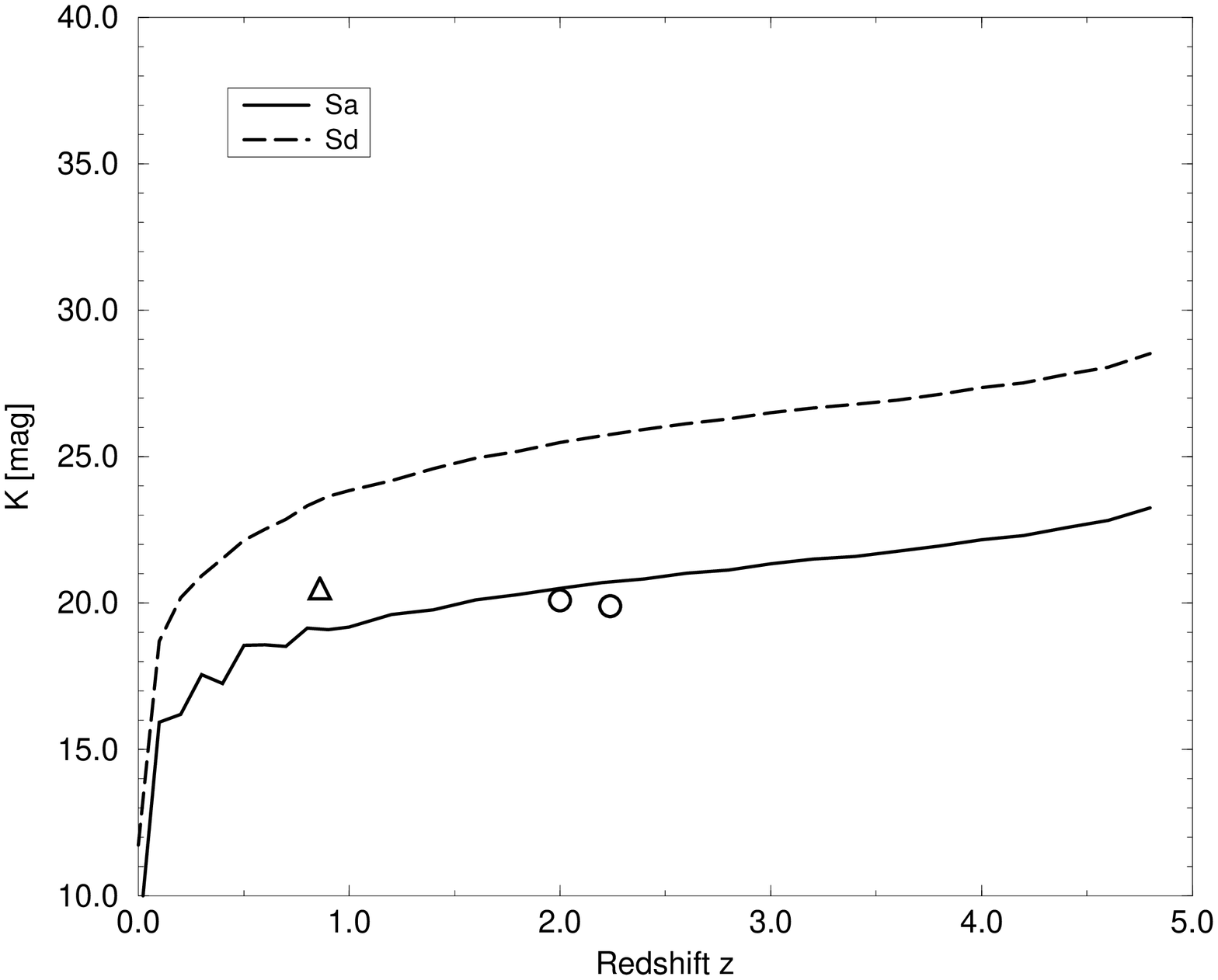,height=5.4cm,angle=270}}
\vspace{-7pt}
\caption{Redshift evolution of apparent magnitudes \R {\bf (2a)} and K {\bf (2b)} for average 
Sa and Sd galaxies. $\Diamond$ galaxies in cluster with DLA absorber (Steidel \etal 1998), 
$\bigtriangleup$ DLA candidate (Steidel \etal 1995), $\bigcirc$ DLA candidates 
(Aragon -- Salamanca \etal 1996).}
\vspace{-10pt}
\end{figure}

\begin{center}
  \begin{tabular}{|l|ccc||ccc|} \hline
	& & Sa & & &Sd & \\ 
	\hline 
	& ~~z$~\sim$ 0.5~~ & & ~~z$~\sim$ 2 -- 3 ~~& ~~z$~\sim$ 0.5~~ & & ~~z$~\sim$ 2 -- 3 ~~\\ 
	\hline 
	~~B~~ & 22.5 & & 24 -- 25 & 25.5 & & 29 -- 30.5  \\ 
      ~~\R~ ~~& 21 & & 24 & 24.5 & & 29 \\
	~~K~~ & 18.5 & & 21.4 -- 22 & 22 & & 26 -- 27 \\
	\hline
  \end{tabular}
\end{center}

In view of these luminosities expected for typical spirals of various types we 
understand why deep surveys could not detect DLA galaxies at z $\sim$ 2 -- 3 
down to \R~ $\sim$ 25.5 (Steidel \etal 1998), and only detected 2 candidates for 10 
DLA systems in the range 1.5 $\lta$ z $\lta$ 2.5 
down to K $\sim$ 21.5 (Aragon -- Salamanca \etal 1996). An average luminosity Sa, if responsible 
for a DLA at z $\lta$ 1.5, would easily have been detected in these surveys. 
DLA candidates identified by Steidel \etal (1994, 1995) 
indeed have M$_{\rm B} \lta -19$ mag typical of (late-type) spirals.

{\bf To conclude:} The change in the DLA galaxy population from progenitors of all 
spiral types at high-z to only late-type spirals (and possibly even LSB galaxies) towards 
low-z, that we derive from the comparison of observed DLA abundances with our chemically consistent 
spiral galaxy models leads to the prediction that, {\bf on average, the low-z DLA galaxies should 
be about as faint in B, \R, and K as the brighter part of the high-z DLA population.} 

%
%
%

\end{document}